 \documentstyle[12pt]{article}
\if@twoside  m
\else
\fi
 1
\font\gt = eufm10 scaled \magstep 1
\font\sr = msbm10 scaled \magstep 1
\font\indsr = msbm10

\def\Tl{${\cal T}^\lambda\ $}
\def\Tla{{\cal T}^\lambda}
\def\Cl{${\cal C}_\lambda\ $}
\def\Cla{{\cal C}_\lambda}
\def\Ml{${\cal M}_\lambda\ $}
\def\Mla{{\cal M}_\lambda}

\def\Mrla{M_\lambda}
\def\tl{$\tau_\lambda\ $}
\def\tla{\tau_\lambda}
\def\Hl{${\cal H}_\lambda\ $}
\def\Hla{{\cal H}_\lambda}
\def\wl{$w_\lambda\ $}
\def\wla{w_\lambda}
\def\el{$e_\lambda\ $}
\def\ela{e_\lambda}

\def\cla{c_\lambda}
\def\Al{$A_\lambda\ $}
\def\Ala{A_\lambda}
\def\Uh{{\cal U}_h}
\def\Aq{{\cal A}_q}
\def\id{{\rm id}}
\def\Ad{{\rm Ad}}
\def\vac{$\langle\cdot\rangle\ $}
\def\vaca{\langle\cdot\rangle}
\def\diag{{\rm diag}\,}
\def\diago{{\rm diag}}
\def\gtg{{\gt g}}
\def\gtk{{\gt k}}
\def\gtb{{\gt b}}
\def\gtn{{\gt n}}
\def\gth{{\gt h}}
\def\gtp{{\gt p}}
\def\Wo{{\cal W}_0}
\def\sgn{{\rm sgn}}
\def\Lin{{\rm Lin}}
\newcommand{\QED}{\mbox{\rule[-1.5pt]{6pt}{10pt}}}
\newcommand{\spav}[1]{\parbox{1mm}{\vspace*{#1}}}

\begin{document}
\begin{titlepage}
\begin{flushright}
CERN-TH.7201/94\\
hep-th/9403114
\end{flushright}
\spav{.1cm}
\begin{center}
{\Large\bf Coherent States for Quantum Compact} \\
{\Large \bf Groups}\\
\spav{1.3cm}\\
{ B. Jur\v co}\\
{\em CERN, Theory Division}\\
{\em CH-1211 Geneva 23, Switzerland}\\
\spav{.5cm}\\
{P. \v S\v tov\'\i\v cek}\\
{\em Department of Mathematics and Doppler Institute}\\
{\em Faculty of Nuclear Science, CTU,
Trojanova~13, 120~00~Prague, Czech Republic}\\

\spav{2cm}\\
{Dedicated to Professor L.D. Faddeev on his 60th birthday}

\end{center}
\spav{.5cm}
\begin{center}
{\bf Abstract}
\end{center}
{\small
 Coherent states  are introduced and their properties are discussed
for all simple quantum compact groups. The multiplicative form of the
canonical element for the quantum double is used to introduce the holomorphic
coordinates on a general quantum dressing orbit and interpret the coherent
state as a holomorphic function on this orbit with values in
the carrier Hilbert space of an irreducible representation of the
corresponding quantized enveloping algebra. Using Gauss decomposition, the
commutation relations for the holomorphic coordinates on the
dressing orbit are derived explicitly and given in a compact R--matrix
formulation
(generalizing this way the $q$--deformed Grassmann and flag manifolds).
The antiholomorphic
realization of the irreducible representations of a compact quantum group
(the analogue of the Borel--Weil construction) are described using the concept
of coherent state. The relation between representation theory and
non--commutative differential geometry is suggested.}

\spav{1.5cm}\\
CERN-TH.7201/94\\
March 1994

\end{titlepage}

\newpage

\setcounter{footnote}{0}
\setcounter{page}{1}
{\thispagestyle{empty}

\section{Introduction}

It is difficult to overestimate the importance of the concept of coherent
states in theoretical and mathematical physics. They found many various
applications in quantum
optics, quantum field theory, quantum statistical mechanics and another
branches of physics as well as in some purely mathematical problems
\cite{Klauder, Perelomov}. The last--named include the Lie group
representations, special functions, automorphic functions, reproducing
kernels etc. In the Lie group representation theory there is a remarkable
relation between the geometry on the coadjoint orbits and the irreducible
representations, which is reflected by the method of orbits (geometric
quantization) due to Kirillov, Kostant and Souriau \cite {Woodhouse}. On the
other hand the concept of coherent states leads naturally to the Berezin's
quantization scheme \cite{Berezin}. The important sources
of both methods are induced representations and the Borel--Weil theory. The
intrinsic relationship between the geometric and Berezin quantization has
been established. There are many papers devoted to this subject (e.g.
\cite{Onofri, Rawnsley} and many others). Recently the coherent states were
used
to construct examples of non--commutative manifolds \cite{G-P}.

The first papers \cite{A-K, Skl}, which can be viewed as those generalizing
the coherent states to the quantum groups appeared even before the formal
birth of quantum groups \cite{Dr}. Number of papers followed subsequently
(\cite{Brano} and many others).
Nevertheless no definition seems to be completely satisfactory. The coherent
states are introduced mainly for the simplest quantum groups
($q$--deformations of the Heisenberg--Weyl, {\gt su}$(2)$ and {\gt su}$(1,1)$
algebras) in a rather straightforward way which does not suggest a proper
generalization to the more general case. Moreover, these states are assumed
to be elements of the representation space for an irreducible representation
of the corresponding quantized enveloping algebra and they do not reflect the
whole underlying Hopf algebra structure.

According to the general philosophy of non--commutative geometry it would be
more natural assume the coherent state as a function on an appropriate
$q$--homogeneous space of the corresponding quantum group (dual to the
quantized enveloping algebra) with values in the representation space.
We hope that such more sophisticated  generalization of the coherent states
method to the case of
quantum groups could be of interest not only for the representation theory
but also for potential applications of quantum groups in physics. Many
important ingredients needed for this generalization are already prepared.
First of all the representation theory of quantum groups \cite{Ji,Lu, Ro}
and the method of induced representations are well developed \cite{P-W}. The
deformations of manifolds playing an important role in the Lie group
representation theory ( such as Schubert cells, flag and Grassmanan manifolds)
are also known \cite{L-R, L-S, T-T, Awata et al, So, Pavel} through
different approaches. Further there is the notion of the quantum dressing
transformation \cite{S-R} the substituent for the coadjoint action from the
classical case, which is important already for classical groups \cite{Se},
has interesting applications in physics \cite{B-B} and is closely related to
the notions of generalized Pontryagin dual and the Iwasawa decomposition
\cite{L-W}. Finally there is also proper definition of the quantum
momentum map \cite{Lu1}. One of the expected results of the coherent
state approach for quantum groups
would be to put all this ingredients together in a natural way.
The second expected result would be a variant of the $q$--generalization
of the Borel--Weil theory which follows more closely to the classical
Borel--Weil construction as this one described in \cite{B-L} for the case of
$U_q(n)$. Finally as in the classical case it is natural to achieve a link
between the representation theory and the non-commutative differential
geometry on quantum groups \cite{Wordc}. We hope to meet this goals in the
present paper.

The present paper prolong some ideas from the papers \cite{J-S, Pavel}, but
now the leading idea is a proper definition of the coherent state for quantum
groups, using all rich structure contained in the quantum double
\cite{Dr, Po-Wo}.

The paper is organized as follows. Section 2 contains a very brief account of
the classical theory. Section 3 has a preliminary character: some basic
notions of the quantum
group theory are recalled. Section 4 adapts to our purposes some well known
results from the representation theory of quantum groups.
Section 5 which contains the definition of coherent state for the compact
quantum group and discusses its basic properties is one of the important
parts of the paper. Here we would like to mention that likewise in the
classical case we can start in the definition (5.1) of the coherent state
$\Gamma$ from any Hopf algebra and any state in the carrier Hilbert space
${\cal H}$ of some its irreducible representation $\tau$ whenever (5.1) does
make sense. Nevertheless the restriction to the quantized universal
enveloping $\Uh$({\gt k}) for {\gt k} compact and the choice of the lowest
(or equivalently the highest) weight state \el are the most relevant for the
rest of the paper. This section also contains a definition of the (quantum)
isotropy subgroup $K_0\subset K$ ($K$ is the spectrum of the Hopf algebra
$\Aq(K)$ dual to the $\Uh$({\gt k})) of \el. Our coherent state can be then
naturally viewed as a function on $q$--homogeneous space $K_0\backslash K$
with values in the representation space \Hl for the lowest weight
representation \tl of $\Uh$({\gt k}) corresponding to the lowest weight
$\lambda$. Section 6 contains a detailed description of the canonical element
$\rho$ (universal R--matrix) of the quantum double (particularly inspired by
\cite{Fronsdal}) which makes possible a more explicit expression for the
coherent state $\Gamma$ and a definition of holomorphic coordinates on a
general quantum dressing orbit. Explicit commutation relations for the
holomorphic coordinates in the R--matrix formulation are derived in
Section 7. They present a compact generalization of the definition relations
for the quantized flag manifold. Section 8 describes the antiholomorphic
realization of the irreducible representation \tl which is most close to the
classical Borel--Weil theory. The presentation of Section 8
can also be, if wished, reinterpreted as a non--commutative version of the
Berezin quantization. Finally in Section 9 we make an attempt to relate the
representation theory to the non--commutative differential geometry, which
as we hope could be helpful for understanding the non--commutative version
of the method of orbits.

Let us make on this place few comments on some points not included in the
paper.

The discussion of Sections 8 and 9 is done using the local coordinates on an
appropriate cell of the dressing orbit. There is no doubt that a
globalization using the quantum Weyl elements is possible. As in the
classical case it has to lead to a "quantization condition" for the quantum
dressing orbit and to an interpretation of the elements of \Hl as
antiholomorphic sections of an appropriate quantum line bundle
\cite{M-B, St}.

There is also no doubt that Section 9 could be formulated purely in terms of
the holomorphic coordinates $z$ and their conjugates $z^{\ast}$. However,
this requires an explicit description of the restriction of the bicovariant
differential calculus on the quantum group $K$ to the quantum homogeneous
space $K_0\backslash K$. An introduction of the partial derivatives
$\partial_{z^{\ast}}$ with respect to the antiholomorphic coordinates
would it make possible to interpret the formula (9.13) expressed only
through coordinates $z^{\ast}$ and  partial derivatives
$\partial_{z^{\ast}}$ as a natural Fock space representation of
$\Uh$({\gt k}).

It is also natural to think about limiting cases of our construction.
The limit $q\rightarrow 1$ gives of course the classical scheme recalled in
Section 2.
Nevertheless as in the classical case \cite{Simon, Berezin} there is a second
type of limit leading to the classical dressing orbits with their natural
Poisson structure. This kind of limit is achieved by using the sequence of
irreducible representations
corresponding to the sequence of lowest weights $n\lambda$. A rescaling of
$q \rightarrow 	q^{1/n}$ and a subsequent limit
$n \rightarrow \infty$ gives the desired result.

\section{The classical scheme}

Let us start from recalling the classical situation
\cite{Perelomov}.
Denote by $G$ a simple and simply connected
complex Lie group and by $K\subset G$ its compact
form. Let \Tl be an irreducible unitary representation of $K$ in \Hl
corresponding to a minimal weight $\lambda$. \Tl extends
unambiguously as
a holomorphic representation of $G$ in \Hl (Weyl unitary trick). Let
$\ela\in\Hla$ be a
normalized
weight vector and set
$$
\Gamma : G\rightarrow\Hla : g\mapsto\Tla(g^{-1})\,\ela\, .
$$
The vector--valued function $\Gamma$ is a coherent state in the sense
of
Perelomov. Denote further by $K_0\subset K$ respectively $P_0\subset
G$ the
isotropy subgroups of the point {\sr C}$\ela\in\,${\sr P}$(\Hla)$.
This means
that there exists a character $\chi$ of $P_0$, unitary on $K_0\subset
P_0$,
such that
$$
\Tla(k)\,\ela =\chi(k)\,\ela\, ,\,\mbox{for}\ k\in P_0\, .
$$
The mapping
$$
\Hla\ni u\mapsto\langle\Gamma(\cdot),u\rangle\in C^\infty(K)
$$
is injective and so one embeds this way \Hl into the vector space of
$\chi$--equivariant functions on $K$. Sending $(g,k)\in K\times K_0$
to
$k^{-1}g\in K$ we get a principal bundle $K\rightarrow K_0\backslash
K$ and
using the 1--dimensional representation $\chi$ one associates to it a
line
bundle over the base space $K_0\backslash K =P_0\backslash G$. Hence $\chi$--
equivariant
functions on $K$ are identified with sections in this line
bundle. Set
$$
\wla:=\langle\ela,\Tla\ela\rangle\in C^{hol}(G)\, .
$$
The function \wl is $\chi$--equivariant on $K$ and thus determines a
trivialization of the line bundle over the cell given by
$\wla(g)\not=0$. The
Gauss decomposition provides a standard way to choose holomorphic
coordinates
$\{z_j\}$ on this cell. Vectors $u$ from \Hl are then represented by
polynomials $\psi_u:=\wla^{\ -1}\langle\Gamma ,u\rangle$ in the
variables
$\{z_j^{\ \ast}\}$ and so the representation \Tl acts in the space of
antiholomorphic functions living on the cell. Finally we also recall
that
every operator $B\in{\rm Lin}(\Hla)$ is represented by its symbol
$\sigma(B)\in
C^a(K_0\backslash K)$ or, this is the same, by a real analytic $K_0$--
invariant
function on $K$,
$$
\sigma(B):=\{ g\mapsto\langle\Gamma(g),B\,\Gamma(g)\rangle\}\, .
$$
The mapping $B\mapsto\sigma(B)$ is injective
\cite{Klauder, Simon}.

The aim of the present paper is to demonstrate that this scheme
applies also
for quantum groups.

\section{Preliminaries, notation}

Let us recall some basic notions related to the duality and the
dressing
transformation for quantum groups
\cite{J-S}.
Concerning the deformation parameter we set $q=e^{-h},\ h>0$. An
important role
plays the duality between the quantum groups $K_q$ and $AN_q$
following
from the Iwasawa
decomposition $G=K\cdot AN$. The deformed enveloping algebra
$\Uh$({\gt k})
is the $\ast$-Hopf algebra dual to $\Aq(K)$. $\Aq(AN)$ is identical to
$\Uh$({\gt k}) as an algebra and opposite as an coalgebra. We note
also that
$\Aq(G)$ is the same Hopf algebra as $\Aq(K)$ but the compact form is
equipped
in addition with the $\ast$-involution. We shall also denote by
$\Uh$({\gt g}) the Hopf algebra $\Uh$({\gt k}) when having forgotten
about
the $\ast$-operation. We denote by $T,\ U$ and $\Lambda$
the vector corepresentations for $\Aq(G),\ \Aq(K)$ and $\Aq(AN)$,
respectively.
The $\ast$-algebras $\Aq(K)$ and $\Aq(AN)$ are defined by the well
known relations
\cite{F-R-T}
$$
RU_1U_2=U_2U_1R,\ U^\ast=U^{-1}\, ,
$$
$$
R\Lambda_1\Lambda_2=\Lambda_2\Lambda_1R,\
\Lambda_1^{\ \ast}R^{-1}\Lambda_2=\Lambda_2R^{-1}\Lambda_1^{\ \ast}\,
,
\eqno(3.1) $$
and for the $B_l,\ C_l$ and $D_l$ series also by
$$
CU^tC^{-1}=U^{-1},\ C\Lambda^tC^{-1}=\Lambda^{-1}\, .
$$
Here $R$ is the standard R-matrix
\cite{Jimbo},
and $C$ is given in
\cite{F-R-T}.
The pairing between $\Aq(AN)$ and $\Aq(K)$ is given by
\cite{J-S}
$$
\langle\Lambda_1;U_2\rangle =R_{21}^{\ -1},\
\langle\Lambda_1^{\ \ast};U_2\rangle =R_{12}^{\ -1}\, .
$$
Let us introduce the canonical element
$$
\rho=\sum x_s\otimes a_s\in\Aq(AN)\otimes\Aq(K)\, ,
$$
with $\{ x_s\}$ and $\{ a_s\}$ being mutually dual bases. Its basic
properties are ($S$ is the antipode, $\Delta$ is the comultiplication)
$$
\rho^\ast=\rho^{-1}=(\id\otimes S)\rho\, ,
$$
$$
(\Delta\otimes\id)\rho=\rho_{23}\rho_{13},\
(\id\otimes\Delta)\rho=\rho_{12}\rho_{13}\, .
\eqno(3.2)$$
Using $\rho$ one defines the dressing transformation as a coaction
$$
{\cal R}: \Aq(AN)\rightarrow\Aq(AN)\otimes\Aq(K):
u\mapsto\rho(u\otimes 1)\rho^{-1}\, .
\eqno(3.3) $$
The identification of the algebras $\Uh$({\gt k}) and $\Aq(AN)$ plays
in this situation the role of the classical momentum mapping.
It is explicitly given by $\Lambda=S(L^+)$ and $\Lambda^{\ast}=L^-$,
where $L^{\pm}$ are the matrices of \cite{F-R-T}.
We note that in the literature one often identifies the dressing
transformation with the quantum adjoint action,
$$
{\rm Ad}_x\, u=\sum x^{(1)}u\, Sx^{(2)},\quad \mbox{with} \quad
\Delta x=\sum x^{(1)}\otimes x^{(2)}\, .
\eqno(3.4) $$
However the both notions are closely related since
$$
(\id\otimes\langle x,\cdot\rangle)\,{\cal R}u={\rm Ad}_x\, u\, ,
\eqno(3.5) $$
where $x\in\Uh$(\gtk) and $u\in\Aq(AN)\equiv\Uh$({\gt k}). The
dressing
transformation can be calculated explicitly on the elements of the
matrix
$\Lambda^\ast\Lambda$,
$$
{\cal R}(\Lambda^\ast\Lambda)=U^\ast\Lambda^\ast\Lambda U\, ,
\eqno(3.6) $$
provided on the RHS one identifies $\Aq(AN)$ with $\Aq(AN)\otimes 1$
and
similarly for $\Aq(K)$.

\section{The "vacuum" functional}

According to the results of Rosso and Lusztig \cite{Ro, Lu},
to every lowest weight $\lambda$ from the weight lattice there is
related
a unique irreducible $\ast$-representation \tl of $\Uh$(\gtk) acting
in
\Hl, dim$\,\Hla<\infty$, and correspondingly a unitary
corepresentation
of $\Aq(K)$, $\Tla=(\tla\otimes\id)\rho\in\Lin(\Hla)\otimes\Aq(K)$.
In what follows, \el stands again for a normalized weight vector.

Let us define the "vacuum" functional \vac on $\Uh$(\gtk),
$$
\langle x\rangle:=\langle\ela,\tla(x)\,\ela\rangle\, .
\eqno(4.1) $$
\proclaim Proposition 4.1.
It holds
$$
\langle x\rangle=\langle x,\wla\rangle\, ,\quad \mbox{where} \quad
\wla:=\langle\ela,\Tla\ela\rangle\in\Aq(K)\, .
\eqno(4.2) $$
This means that \vac if viewed as an element from $\Aq(K)$, the
dual space to $\Uh$({\gt k}), is equal to \wl.

\noindent {\em Proof.}
One can verify (4.2) easily using the identity
$$
(\id\otimes\langle x,\cdot\rangle)\rho=x,\
x\in\Uh(\hbox{\gtk})\equiv\Aq(AN)\, .\quad\QED
$$
Let us note at this place that, likewise in the classical case,
\proclaim Proposition 4.2.
It holds
$$
w_{\lambda_1+\lambda_2}=w_{\lambda_1}w_{\lambda_2}=
w_{\lambda_2}w_{\lambda_1}\, ,
\eqno(4.3) $$
and so it is enough to determine \wl only for the fundamental weights
$\lambda=\omega_j$.
Furthermore ($\varepsilon$ is the counit),
$$
S\,\wla=\wla^{\ \ast}, \quad \varepsilon(\wla)=1\, .
\eqno(4.4) $$

\noindent {\em Proof.} To see (4.3) it suffices to observe that
${\cal H}_{\lambda_1+\lambda_2}$ can be identified with the cyclic
submodule $\cal M$ in ${\cal H}_{\lambda_1}\otimes{\cal H}_{\lambda_2}$
corresponding to the cyclic vector
$e_{\lambda_1}\otimes e_{\lambda_2}$ with respect to the representation
$(\tau_{\lambda_1}\otimes\tau_{\lambda_2})\circ\Delta$. Since
$$
{\cal T}^{\lambda_1+\lambda_2}=
{\cal T}^{\lambda_2}_{\quad 23}{\cal T}^{\lambda_1}_{\quad 13}\vert
{\cal M}\, ,
$$
we have
\begin{eqnarray*}
w_{\lambda_1+\lambda_2} & = &
\langle e_{\lambda_1}\otimes e_{\lambda_2},{\cal
T}^{\lambda_1+\lambda_2}
e_{\lambda_1}\otimes e_{\lambda_2}\rangle\cr\cr
& = & \langle e_{\lambda_2},{\cal T}^{\lambda_2}e_{\lambda_2}\rangle
\langle e_{\lambda_1},{\cal T}^{\lambda_1}e_{\lambda_1}\rangle\cr\cr
& = & w_{\lambda_2}w_{\lambda_1}\, .\quad\QED \cr\cr
\end{eqnarray*}

Using the identification $\Uh$({\gt k})$\equiv\Aq(AN)$ one can also
describe the "vacuum" functional in the following way. It holds
$$
\tla(\Lambda)\,\ela =\Ala\ela\, ,
\eqno(4.5) $$
where \Al is a positive diagonal matrix fulfilling the $R\Ala\Ala$--
equation
and possibly also $C\Ala^{\ t}C^{-1}=\Ala^{\ -1}$. Besides, the
relation (3.1)
enables one to define a normal ordering on $\Aq(AN)$ by requiring the
elements
of the matrix $\Lambda^\ast$ to stand to the left and those of the
matrix
$\Lambda$ to stand to the right. It doesn't matter that this ordering
prescription is not quite unambiguous since the subalgebras generated
by the
entries of $\Lambda^\ast$ and $\Lambda$, respectively, intersect in
the Cartan
elements. We have
$$
\langle 1\rangle=1,\ \
\langle\Lambda^\ast\rangle=\langle\Lambda\rangle=\Ala\, ,
\eqno(4.6) $$
and
$$
\langle x_{i_1}\dots x_{i_k}\rangle=
\langle x_{i_1}\rangle\dots\langle x_{i_k}\rangle\, ,
$$
provided the product $x_{i_1}\dots x_{i_k}$ is normally ordered.

\section{The quantum coherent state}

The following definition is very analogous to the classical case and is
crucial for the rest of the paper.
\proclaim Definition 5.1.
We define the quantum coherent state as the element
\begin{eqnarray*}
\Gamma & := & (\Tla)^{-1}(\ela\otimes 1) & {\rm (5.1)}\cr\cr
& = & (\tla\otimes S)\rho\cdot(\ela\otimes 1)\in\Hla\otimes\Aq(K)\, .&\cr\cr
\end{eqnarray*}

\noindent $\Gamma$ should be interpreted as a quantum function on $K$ with
values in
\Hl . Thus one can relate to every vector $u\in\Hla$ a quantum
function on
$K$,
$$
u\mapsto\langle\Gamma,u\rangle
:=(\langle\ela,(\cdot)u\rangle\otimes\id)\Tla
\in\Aq(K)\, .
\eqno(5.2) $$
Furthermore, the operators in \Hl can be again represented by their
symbols,
$$
\sigma :\Lin(\Hla)\to\Aq(K): B\mapsto\langle\Gamma,B\Gamma\rangle\, .
\eqno(5.3) $$

\proclaim Proposition 5.2.
The mapping $\sigma$ is injective.

\noindent{\em Proof.}
The proof goes through as in the classical case \cite{Klauder, Simon}.
Let us sketch it. $\sigma(B)=0$ means that
$$
\langle\ela,\Tla B\, S(\Tla)\,\ela\rangle=0\, .
$$
Applying k--times the comultiplication to the LHS, pairing with the elements
$X^-_{i_k}\dots X^-_{i_1}$ and using the fact that \el is the lowest
weight vector and that $(X^\pm_i)^\ast=X^\mp_i$ we obtain
$$
\langle\tla(X^+_{i_1})\dots\tla(X^+_{i_k})\,\ela,B\ela\rangle=0\, .
$$
Since the vectors $\tla(X^+_{i_1})\dots\tla(X^+_{i_1})\,\ela$ span
\Hl it
follows that $B\ela=0$. Applying instead the comultiplication
$(k+1)$--times one finds that the same argument is valid also
provided $B$ is replaced $\Tla B\, S(\Tla)\in\Lin(\Hla)\otimes\Aq(K)$
and so $B\, S(\Tla)\,\ela=0$. The same reasoning as above gives $B=0$. \QED

The symbol can be extended naturally as a mapping from $\Uh$({\gt
k})$\equiv
\Aq(AN)$ to $\Aq(K)$ by putting
$$
\sigma=(\vaca\otimes\id)\circ{\cal R}\, .
\eqno(5.3) $$
Now we are ready to define the isotropy subgroup as a $\ast$-Hopf
algebra
$\Aq(K_0)$ with the vector representation $U_0$ and the projection
("restriction morphism") $p_0:\Aq(K)\to\Aq(K_0),\ p_0(U)=U_0$. We
require
$$
(\vaca\otimes p_0)\circ{\cal R}=\vaca\, 1
\eqno(5.4) $$
as a morphism from $\Aq(AN)$ to {\sr
C}$\otimes\Aq(K_0)\equiv\Aq(K_0)$, i.e.,
$$
p_0(\sigma(Y))=\langle Y\rangle\, 1,\quad \mbox{for} \quad Y\in\Aq(AN)\, .
\eqno(5.5) $$
According to (3.6) and (4.6),
$$
\sigma(\Lambda^\ast\Lambda)=U^{-1}\Ala^{\ 2}U\, .
\eqno(5.6) $$
Consequently, in addition to the equations
$$
RU_{01}U_{02}=U_{02}U_{01}R,\ U_0^{\ \ast}=U_0^{\ -1}\, ,
\eqno(5.7) $$
and for $B_l,\ C_l,\ D_l$ also $CU_0^{\ t}C^{-1}=U_0^{\ -1}$,
$U_0$ should fulfill
$$
U_0^{\ -1}\Ala^{\ 2}U_0=\Ala^{\ 2}\, .
\eqno(5.8) $$
The condition (5.8) is formally the same as in the classical case. In
fact it
amounts in annulation of some entries of the matrix $U$ when taking
the
projection $p_0(U)=U_0$. For the enveloping algebra $\Uh$({\gt
k}$_0$) this
means that there exits a subset $\Pi_0$ of the set of simple roots
$\Pi$ so
that $\Uh$({\gt k}$_0$) is generated by all Cartan elements $H_i$ and
only
by those elements $X^\pm_i$ for which $\alpha_i\in\Pi_0$.

Thus on the dual level we have an injection $\Uh$({\gt
k}$_0)\hookrightarrow
\Uh$({\gt k}). An element $X$ from $\Uh$({\gt k}) belongs to
$\Uh$({\gt k}$_0$) if and only if
$$
\langle X,f\,\sigma(Y)\, g\rangle=\langle\Delta X,f\otimes g\rangle
\langle Y\rangle
\eqno(5.9) $$
holds for every $Y\in\Uh$({\gt k}) and $f,g\in\Aq(K)$. Letting
$f=g=1$ we have
(c.f. (3.5))
$$
\langle\Ad_X\, Y\rangle=\langle X,\sigma(Y)\rangle=\varepsilon(X)\,
\langle Y\rangle\, .
\eqno(5.10) $$
Let us substitute the elements $H_i$ and $X^+_i$ for $Y$ in (5.10).
Using $\tla(H_i)\,\ela=\lambda(H_i)\,\ela$ and $\tla(X^-_i)\,\ela=0$ we
find that (5.10) is true for all Cartan elements $H_i$ and only for those
elements $X^+_i$ which fulfill
$$
\langle Y\,
X^+_i\rangle=\langle\tla(Y^\ast)\,\ela,\tla(X^+_i)\,\ela\rangle
=0\, .
$$
Putting $Y^\ast=X^+_{i_1}\dots X^+_{i_k}$ we conclude that the
condition on the
subset $\Pi_0\subset\Pi$ is:
$$
\alpha_i\in\Pi_0\quad \mbox{iff} \quad\tla(X^+_i)\,\ela=0\, .
\eqno(5.11) $$

It follows immediately that there exists a character $\chi$ on
$\Uh$({\gt k}$_0$) such that
$$
\tla(X)\,\ela=\chi(X)\,\ela,\quad \mbox{for} \quad X\in\Uh(\hbox{\gtk}_0)\, .
\eqno(5.12) $$
Pairing the both sides with \el one finds that $\chi(\cdot)$ is the
restriction of the "vacuum" functional \vac. Considering $\chi$ as an
element
from $\Aq(K_0)$ we deduce that
$$
\chi=p_0(\wla)\quad \mbox{and} \quad\Delta\chi=\chi\otimes\chi\, .
\eqno(5.13) $$
Moreover, using (5.13), (4.4) and the relation
$m\circ(S\otimes\id)\circ\Delta=
\varepsilon$ we have
$$
S\chi=\chi^\ast=\chi^{-1}\, .
\eqno(5.14) $$

Let us now check the equivariance property. First note that from (5.3)
and the relation $(\id\otimes\Delta){\cal R}=({\cal
R}\otimes\id){\cal R}$ it follows
\proclaim Proposition 5.3.
$$
\Delta\circ\sigma=(\sigma\otimes\id)\circ{\cal R}\, ,
\eqno(5.15) $$
and hence, by (5.5),
$$
(p_0\otimes\id)\,\Delta\,\sigma(Y)=1\otimes\sigma(Y)\, .
\eqno(5.16) $$
This means that every symbol $\sigma(Y)\in\Aq(K)$ is left $K_0$--
invariant, i.e., $\sigma(Y)\in\Aq(K_0\backslash K)$.

Concerning the equivariance of the coherent state itself we have
\proclaim Proposition 5.4.
It holds
$$
(p_0\otimes\id)\,\Delta\langle\Gamma,u\rangle  =
\chi\otimes\langle\Gamma,u\rangle\, .
\eqno(5.17) $$
Particularly, putting $u=\ela$, we have
$$
(p_0\otimes\id)\,\Delta\wla=\chi\otimes\wla\, .
\eqno(5.18) $$
So the quantum function $\wla^{\ -1}\langle\Gamma,u\rangle$ is left
$K_0$--invariant and belongs to some completion of the algebra
$\Aq(K_0\backslash K)$ since we admit \wl to be invertible.

\noindent{\em Proof.}
First note that (5.12) can be rewritten dually as
$$
(\id\otimes p_0)\Tla\cdot(\ela\otimes 1)=\ela\otimes\chi\, .
$$
Hence, using the unitarity of \Tl and $\chi$, we have for any $u\in\Hla$,
$$
(\langle\ela,(\cdot)u\rangle\otimes
p_0)\Tla=\langle\ela,u\rangle\,\chi\, .
$$
It follows that
\begin{eqnarray*}
(p_0\otimes\id)\,\Delta\langle\Gamma,u\rangle & = &
(\langle\ela,(\cdot)\rangle\otimes p_0\otimes\id)\Tla_{\ 12}\Tla_{\ 13}
& \cr\cr
& = & (\langle\ela,(\cdot)u\rangle\chi\otimes\id\rangle\Tla & \cr\cr
& = & \chi\otimes\langle\Gamma,u\rangle\, .\ \QED  & \cr\cr
\end{eqnarray*}

\section{Canonical element for the double}

The complex structure on the quantized homogeneous space $K_0\backslash K$ is
introduced the same way as in the classical case. Namely, the subalgebra of
$\Aq(K_0\backslash K)$ consisting of holomorphic functions coincides
with $\Aq(P_0\backslash G)$. Here $\Aq(P_0)$ is the Hopf algebra dual to
$\Uh$({\gt p}$_0$), the Hopf subalgebra in $\Uh$({\gt g}) generated
by all $H_i,\ X^-_i$ and those $X^+_i$ for which $\alpha_i\in\Pi_0$. Since
the condition (5.12)
clearly extends to all $X\in\Uh$({\gt p}$_0$) it is easy to verify
that for every $u\in\Hla$, $\langle u,\Gamma\rangle\,(\wla^{\ \ast})^{-1}$
is a holomorphic quantum function. Thus one can represent vectors
from \Hl by antiholomorphic functions,
$$
u\mapsto \psi_u:=\wla^{\ -1}\langle\Gamma,u\rangle\, .
\eqno(6.1) $$
This mapping is injective as one can show using the same reasoning as
in the case of the symbol (Sec. 5). It is desirable to introduce quantum
(non-commutative) local holomorphic coordinates $z_j$ on
$K_0\backslash K$ and
consequently to express $\psi_u=\psi_u(z_j^{\ \ast})$ as an polynom in
$z_j^{\ \ast}$. To this end we shall employ the Gauss decomposition.

Denote by {\gt b}$_\pm\subset\,${\gt g} the Borel subalgebras and by
{\gt h}$=\,${\gt b}$_+\cap\,${\gt b}$_-$ the Cartan subalgebra. It is
known that the Hopf algebras $\Uh(${\gt b}$_+)$ and $\Uh(${\gt b}$_-
)^{op\Delta}$
are mutually dual and that the dual quantum double for $\Uh(${\gt b}$_+)$ can
be identified as an algebra with  $\Uh$({\gt g})$\otimes\Uh$({\gt h}). To
have this identification also for the coalgebras one has to twist the
comultiplication in $\Uh$({\gt g})$\otimes\Uh$({\gt h}) using the element
$\exp(\sum H_i^0\otimes H_i^0)$ with $\{ H_i^0\}$ being any orthonormal basis
in {\gt h} \cite{S-R}.
According to the terminology we have adopted here the dual quantum double
means twisted multiplication while the quantum double means twisted
comultiplication. Thus on the dual level one obtains for the
corresponding
algebras of quantum functions,
$$
\Aq(B_-)\otimes\Aq(B_+)\simeq\Aq(G)\otimes_{\rm twist}\Aq(H)\, .
\eqno(6.2) $$
The vector corepresentations $L^{(\pm)}$ and $J$ of the quantum groups
$(B_\pm)_q$ and $H_q$, respectively, fulfill the corresponding
$RXX$--equations
and possibly also the deformed orthogonality condition. For a proper choice
of the set $\Pi$ of simple roots, $L^{(\pm)}$ is upper (lower) triangular and
$J$ is diagonal. The isomorphism in (6.2) is given by
$$
T\equiv T\otimes 1=L^{(-)}\dot\otimes L^{(+)},\
J\equiv1\otimes J=(\diag L^{(-)})^{-1}\diag L^{(+)}\, ,
\eqno(6.3) $$
and the twisted comultiplication on the RHS is determined by
$$
\diago(R)\, T_1J_2=J_2T_1\diago(R)\, .
\eqno(6.4) $$

This structure has turned out to be very helpful in construction of
the universal R-matrix $R^u\in\Uh$(\gtg)$\otimes\Uh$(\gtg).
\cite{kirillov-Reshetikhin,Levendorskii-Soibelman}
First by fixing a maximal Weyl element one orders the set $\Delta^+$
of positive roots as $(\beta_1,\dots,\beta_d),\ d=|\Delta^+|$. To each
root $\beta_j$ there are related elements $E(j)\in\Uh$(\gtb$_+)$ and
$F(j)\in\Uh$(\gtb$_-)$ so that the elements
$$
E(d)^{n_d}\dots E(1)^{n_1}H_l^{\ m_l}\dots H_1^{\ m_1}\, ,
\eqno(6.5) $$
$n_i,\ m_i\in\,${\sr Z}$_+$, form a basis in $\Uh$(\gtb$_+)$. The vectors
$H_i$ can be replaced by any elements forming a basis in \gth and a similar
assertion is valid also for $\Uh$(\gtb$_-)$. In the limit
$h\downarrow 0$
the elements $E(j)$ and $F(j)$ become the root vectors
$X_{\beta_j}\in\,$\gtn$_+$ and $X_{-\beta_j}\in\,$\gtn$_-$,
respectively. We recall that the universal R-matrix can be written in the form
\cite{khoroskhin-Tolstoy}
$$
R^u=\exp_{q_d}\bigl(\mu_d\, F(d)\otimes E(d)\bigr)\dots
\exp_{q_1}\bigl(\mu_1\, F(1)\otimes E(1)\bigr)\, \exp(\kappa)\, ,
\eqno(6.6) $$
where $\exp_q$ are the q-deformed exponential functions, $\mu_j$ are
some coefficients depending on the parameter $h$ and $\kappa$ is some
element from $\Uh$(\gth)$\otimes\Uh$(\gth).

Equipped with this knowledge we are able to reveal the structure of
the canonical element for the double $\Aq(AN)\otimes\Aq(K)$. We make use
of the fact that
$\Aq(AN)\simeq\Uh$(\gtg)$^{op\Delta}$ is a factoralgebra of
$\Uh$(\gtb$_-)^{op\Delta}\otimes_{\rm twist}\Uh$(\gtb$_+)^{op\Delta}$ and
$\Aq(K)\simeq\Uh$(\gtg)$^\ast$ is a subalgebra in
$\Uh$(\gtb$_+)^{op\cdot}\otimes\Uh$(\gtb$_-)^{op\Delta}$. The canonical
element $\tilde\rho$ in
$$
\Bigl(\Uh(\hbox{\gtb}_-)^{op\Delta}\otimes_{\rm twist}
\Uh(\hbox{\gtb}_+)^{op\Delta}\Bigr)\otimes\Bigl(
\Uh(\hbox{\gtb}_+)^{op\cdot}\otimes\Uh(\hbox{\gtb}_-)^{op\Delta}\Bigr)
\eqno(6.7) $$
can be decomposed as follows
\cite{FRT}
\begin{eqnarray*}
\tilde\rho & = & \sum (e_j\otimes e^k)\otimes(f^j\otimes f_k) & \cr\cr
& = & \sum (e_j\otimes 1\otimes f^j\otimes 1)\cdot
(1\otimes e^k\otimes 1\otimes f_k) & (6.8) \cr\cr
& = & \tilde R_{13}\tilde R'_{24}\, . & \cr\cr
\end{eqnarray*}
Here $\{e_j\},\ \{e^k\},\ \{f^j\}$ and $\{f_k\}$ stand for bases in the
corresponding factors, $\{e_j\}$ and $\{f^j\}$ are dual and the same  is
assumed about $\{e^k\}$ and $\{f_k\}$, the dot in the third member of
equalities (6.8) indicates multiplication in the double and $\tilde R'$ is
obtained from $\tilde R$ by reversing the order of multiplication. To express
$\rho$ we shall use again bases of the type (6.5). In our notation
the elements
$F(j),\ E(j),\ \tilde E(j)$ and $\tilde F(j)$ belong in this order to the
individual factors in (6.7). Factorizing off the redundant Cartan
elements we obtain finally
\proclaim Proposition 6.1.
The canonical element for the quantum double $\Aq(AN)\otimes\Aq(K)$
has the form
\begin{eqnarray*}
\rho & = & \exp_{q_d}\bigl(\mu_d\, F(d)\otimes\tilde E(d)\bigr)\dots
\exp_{q_1}\bigl(\mu_1\, F(1)\otimes\tilde E(1)\bigr)\, \exp(\kappa)
& (6.9) \cr\cr
& & \times\exp_{q_1}\bigl(\mu_1\, E(1)\otimes\tilde F(1)\bigr)\dots
\exp_{q_d}\bigl(\mu_d\, E(d)\otimes\tilde F(d)\bigr)\, . & \cr\cr
\end{eqnarray*}

To proceed further in this analysis we note that the maximal Weyl element can
be chosen so that there exists $p\in\,${\sr Z}$_+$, $p\leq d$, such that
the vectors $X_{-\beta_1},\dots, X_{-\beta_d}$, $H_1,\dots, H_l$,
$X_{\beta_1},\dots,X_{\beta_p}$
form a basis of \gtp$_0$. Then $X_{\beta_{p+1}},\dots,X_{\beta_d}$
form a basis of a nilpotent subalgebra \gtn$_0$ and
\gtg$=$\gtp$_0\oplus$\gtn$_0$.
Notice that in the generic case $\Pi_0=\emptyset$ and hence $p=0,$
\gtp$_0=\,$\gtb$_-$
and \gtn$_0=\,$\gtn$_+$. This means that all elements $F(j)$ belong to
$\Uh$(\gtp$_0)$ while $E(j)$ belongs to $\Uh$(\gtp$_0)$ only for
$j=1,\dots,p$. Consequently,
\begin{eqnarray*}
\tla(F(j))\,\ela & = & 0\, ,\quad \mbox{for} \ j=1,\dots,d, & (6.10) \cr\cr
\tla(E(j))\,\ela & = & 0\, ,\quad \mbox{for} \ j=1,\dots,p,\, . & \cr\cr
\end{eqnarray*}
Let $\tau$ designate the irreducible representation of $\Uh$(\gtg)
corresponding to the vector corepresentation $T$ of $\Aq(G),\
T=(\tau\otimes\id)\rho$. We have
\proclaim Corollary 6.2.
$T$ can be written as a product,
\begin{eqnarray*}
T & = & \Lambda_{(-)}Z\, ,\quad \mbox{where} & (6.11) \cr\cr
\Lambda_{(-)} & = & (\tau\otimes\id)\,
\exp_{q_d}\bigl(\mu_d\, F(d)\otimes\tilde E(d)\bigr)\dots
\exp_{q_1}\bigl(\mu_1\, F(1)\otimes\tilde E(1)\bigr)\, \exp(\kappa) &
\cr\cr
& & \times\exp_{q_1}\bigl(\mu_1\, E(1)\otimes\tilde F(1)\bigr)\dots
\exp_{q_p}\bigl(\mu_p\, E(p)\otimes\tilde F(p)\bigr)\, , & \cr\cr
Z & = & (\tau\otimes\id)\,
\exp_{q_{p+1}}\bigl(\mu_{p+1}\, E(p+1)\otimes\tilde F(p+1)\bigr)\dots
\exp_{q_d}\bigl(\mu_d\, E(d)\otimes\tilde F(d)\bigr)\, . & \cr\cr
\end{eqnarray*}
The matrix $\Lambda_{(-)}$ is block lower triangular, $Z$ is block upper
triangular and the blocks on the diagonal of $Z$ are unit matrices.

\noindent{\em Remark.}
The splitting into the blocks is determined by decomposition of \gtg$_0=$
complexification of \gtk$_0$ into the direct sum of simple subalgebras and
an Abelian subalgebra and it will be described more explicitly in the next
section. In the generic case of $\Pi_0=\emptyset$,
\gtg$_0=\,$\gth\ and the matrices $\Lambda_{(-)}$ and $Z$ are simply
lower and upper triangular.

Notice that the entries of $Z$ are expressed as polynomials in
$d-p={\rm dim}_{\hbox{\indsr C}}(P_0\backslash G)$ noncommutative variables
$\tilde F(p+1),\dots,\tilde F(d)$ and can be considered as local
holomorphic coordinates on the orbit. Next we are going to derive
explicit commutation relations for them.

Recalling the definition (5.1) of the coherent state $\Gamma$ and
using the relations (6.9), (6.10), we obtain
$$
\Gamma=\exp^{-1}_{q_d}\bigl(\mu_d\,\tla(E(d))\otimes\tilde
F(d)\bigr)\dots
\exp^{-1}_{q_{p+1}}\bigl(\mu_{p+1}\,\tla(E(p+1))\otimes\tilde
F(p+1)\bigr)
\cdot(\ela\otimes\wla^{\ \ast})\, ,
\eqno(6.12) $$
since
$$
\wla^{\ \ast}=(\langle\ela,\tla(\cdot)\ela\rangle\otimes\id)\,\rho^{-1}=
\exp\Bigl((\langle\ela,\tla(\cdot)\ela\rangle\otimes\id)\,\kappa\Bigr)
$$
Thus we find again that, for every $u\in\Hla$, $\psi_u$ given by
(6.1) is an
antiholomorphic quantum function and should be expressible in the
variables $z^\ast$.

\section{Quantum holomorphic coordinates on a ge\-neral dressing orbit}

We start from the decomposition $T=\Lambda_{(-)}Z$. Let now $p_0$ stand for
the "restriction" morphism $\Aq(G)\to\Aq(P_0)$. First we shall verify
that the entries of $Z$ are left $P_0$--invariant quantum functions. We have
$$
(p_0\otimes\id)\,\Delta T=(p_0\otimes\id)\,\Delta\Lambda_{(-)}\cdot
(p_0\otimes\id)\,\Delta Z\, .
$$
At the same time,
$$(p_0\otimes\id)\,\Delta T=p_0(T)\dot\otimes T=
(p_0(T)\dot\otimes\Lambda_{(-)})({\bf 1}\dot\otimes Z)\, .
$$
Since the decomposition into a product of block lower triangular and block
upper triangular matrices, the latter having unit diagonal blocks, is
unambiguous we find by comparing that
$$
(p_0\otimes\id)\,\Delta Z={\bf 1}\dot\otimes Z\, .
\eqno(7.1) $$

To derive commutation relations for the matrix elements of $Z$ one can again
employ the Gauss decomposition. This time we have in mind the isomorphism
(6.2), (6.3). We are going to enumerate the matrix elements in the vector
representation by weights. This is possible since for all four principal
series $A,\ B,\ C,\ D$, the weights of the vector representations are simple.
Every weight belongs either to the Weyl group orbit of the corresponding
fundamental weight or is zero (only for the series $B$). We shall use
the standard ordering on the set of weights: $\sigma>\nu$ iff
$\sigma\not=\nu$
and $\sigma-\nu=\sum m_i\alpha_i$, with $m_i\in\,${\sr Z}$_+\
(0\in\,${\sr Z}$_+)$. Set
$$
{\cal W}_0=\bigoplus_{\alpha_i\in\Pi_0}\,\hbox{\sr Z}_+\alpha_i\, .
\eqno(7.2) $$
We shall write simply $L=(L_{\sigma\nu})$ instead of $L^{(+)}$. Thus
$L_{\sigma\nu}=0$ whenever $\sigma<\nu$ (pay attention, the ordering on
weights is reversed in comparison with the standard enumeration of weights).
Further we introduce a matrix $A$ by
\begin{eqnarray*}
A_{\sigma\nu} & = & L_{\sigma\nu}\, ,\quad  \mbox{if} \ \sigma-\nu\in{\cal
W}_0\, ,
& (7.3) \cr\cr
& = & 0\, ,\quad\quad   \mbox{otherwise}. & \cr\cr
\end{eqnarray*}
Comparing (6.3) and (6.11) we obtain
$$
Z=A^{-1}L\, .
\eqno(7.4) $$

Next we recall a useful property of the R-matrix. Namely,
$R_{\sigma\tau,\mu\nu}\not=0$ implies $\sigma-\mu=\nu-\tau,\
\sigma\leq\mu,\
\tau\geq\nu$, and one of the following three possibilities happens:\\
$(i)\ \sigma=\mu,\ \tau=\nu\, ,$\\
$(ii)\ \sigma=\nu<\tau=\mu\, ,$\\
$(iii)\ \sigma=-\tau<\mu=-\nu\, .$\\

We continue by deriving some auxiliary relations. The first one is
\proclaim Lemma 7.1.
It holds
$$
\Delta A=A\dot\otimes A\, ,\quad \mbox{in} \quad \Aq(B_+)\, ,
\eqno(7.5) $$
and consequently
$$
R\, A_1A_2=A_2A_1R\, .
\eqno(7.6) $$

\noindent{\em Proof.} In the equality
$$
\Delta L_{\sigma\nu}=\sum_\xi L_{\sigma\xi}\otimes L_{\xi\sigma}\, ,
$$
the nonzero summands should fulfill $\sigma\geq\xi\geq\nu$. To obtain
(7.5) it is enough to notice that then
$\sigma-\nu\in{\cal W}_0$ implies $\sigma-\xi,\,\xi-\nu\in{\cal W}_0$.

The relation (7.6) is the same as
$$
\langle Y,R\, A_1A_2-A_2A_1R\rangle=0\, ,\quad \mbox{for all}\quad
Y\in\Uh(\hbox{\gtb})_+ .
$$
The last equality can be deduced from the following facts. This
relation is valid provided $A$ is replaced by $L$. Clearly
$\langle X_i^+,A\rangle=0$
whenever $\alpha_i\not\in\Pi_0$ and so
\begin{center}
$\langle Y_1X_i^+Y_2,A\rangle=0\, ,\quad \mbox{for}\ \alpha_i\not\in\Pi_0\
\mbox{and
any}\ Y_1,Y_2\in\Uh$({\gtb}$_+)$ .
\end{center}
Finally,
\begin{eqnarray*}
\langle H_i,A\rangle & = & \langle H_i,L\rangle\, ,\quad \mbox{for all}\
i,\cr\cr
\langle X_i^+,A\rangle & = & \langle X_i^+,L\rangle\, ,\quad
\mbox{provided}\ \alpha_i\in\Pi_0\, .\ \QED \cr\cr
\end{eqnarray*}

By annulating some entries of the R-matrix we define another matrix
$Q=Q_{12}$,
\begin{eqnarray*}
Q_{\sigma\tau,\mu\nu} & = & R_{\sigma\tau,\mu\nu}\, ,\quad
\mbox{provided}\ \tau-\nu=\mu-\sigma\in{\cal W}_0\, , & (7.7) \cr\cr
& = & 0\, ,\qquad\quad \mbox{otherwise}. & \cr\cr
\end{eqnarray*}
\proclaim Lemma 7.2.
It holds
$$
Q\, L_1A_2=A_2L_1Q\,
\eqno(7.8) $$
and
$$
Q\, A_1A_2=A_2A_1Q\, .
\eqno(7.9) $$

\noindent{\em Proof.} To show (7.8) assume in the equality
$$
\sum_{\xi\eta} R_{\sigma\tau,\xi\eta}\, L_{\xi\mu}\, L_{\eta\nu}=
\sum_{\xi\eta} L_{\tau\eta}\, L_{\sigma\xi}\, R_{\xi\eta,\mu\nu}\, ,
$$
that $\tau-\nu\in{\cal W}_0$. The nonzero summands on the both sides
should fulfill $\tau\geq\eta\geq\nu$ whence
$\tau-\eta,\,\eta-\nu\in{\cal W}_0$.
Thus we obtain
$$
\sum_{\xi\eta} Q_{\sigma\tau,\xi\eta}\, L_{\xi\mu}\, A_{\eta\nu}=
\sum_{\xi\eta} A_{\tau\eta}\, L_{\sigma\xi}\, Q_{\xi\eta,\mu\nu}\, ,
\eqno(7.10) $$
It remains to verify validity of (7.10) also for
$\tau-\nu\not\in{\cal W}_0$.
Again, the nonzero summands on the both sides of (7.10) should satisfy
$\tau-\eta,\,\eta-\nu\in{\cal W}_0$. But ${\cal W}_0$ is additive and
so this can never happen.

Let us show (7.9). Assume in (7.10) that $\mu\geq\sigma$. The nonzero
summands on the LHS should fulfill $\xi-\sigma\in{\cal W}_0$ and
$\xi\geq\mu\geq\sigma$
whence $\xi-\mu\in{\cal W}_0$. Analogously for the RHS we have
$\mu-\xi\in{\cal W}_0$ and $\mu\geq\sigma\geq\xi$ whence $\sigma-
\xi\in\Wo$.
Thus we obtain in this case
$$
\sum_{\xi\eta} Q_{\sigma\tau,\xi\eta}\, A_{\xi\mu}\, A_{\eta\nu}=
\sum_{\xi\eta} A_{\tau\eta}\, A_{\sigma\xi}\, Q_{\xi\eta,\mu\nu}\, .
\eqno(7.11) $$
Next assume in (7.10) that $\sigma-\mu\in\Wo$. The nonzero summands on
the LHS should fulfill $\xi-\sigma\in\Wo$ whence, owing to the additivity,
$\xi-\mu\in\Wo$. Analogously for the RHS we have $\mu-\xi\in\Wo$
and hence $\sigma-\xi\in\Wo$. Also in
this case we arrive at (7.11). It remains to verify (7.11) for
$\sigma>\mu$ but
$\sigma-\mu\not\in\Wo$. Now the nonzero summands on the LHS of (7.11)
should fulfill $ \xi-\sigma,\,\xi-\mu\in\Wo$. But this can never happen
since then
$\mu<\sigma\leq\xi$ and $\sigma-\mu$ would belong to $\Wo$.
Analogously on
the RHS, it never happens that, at the same time, $\mu-\xi$ and
$\sigma-\xi$ belong to $\Wo$. \QED

The final relation we shall need is
\proclaim Lemma 7.3.
It holds
$$
A_2^{\ -1}Z_1A_2=Q^{-1}Z_1Q\, .
\eqno(7.12) $$

\noindent{\em Proof.}
One can verify (7.12) by using in (7.8) the substitution $L=AZ$ and the
equality (7.9),
$$
A_2A_1Z_1Q=Q\, A_1A_2A_2^{\ -1}Z_1A_2=A_2A_1QA_2^{\ -1}Z_1A_2\, .\ \QED
$$

Now we are able to state the desired commutation relation.
\proclaim Proposition 7.4.
The matrix $Z$ obeys the equality
$$
R\, Q_{12}^{\ -1}Z_1Q_{12}Z_2=Q_{21}^{\ -1}Z_2Q_{21}Z_1R\, .
\eqno(7.13) $$

\noindent{\em Proof.}
To prove (7.13) use the substitution $L=AZ$ in the $RLL$--equation,
$$
R\, A_1A_2(A_2^{\ -1}Z_1A_2)Z_2=A_2A_1(A_1^{\ -1}Z_2A_1)Z_1R\, ,
$$
and apply (7.6) and (7.12),
$$
A_2A_1R\, Q_{12}^{\ -1}Z_1Q_{12}Z_2=
A_2A_1Q_{21}^{\ -1}Z_2Q_{21}Z_1R\, .\ \QED
$$

This result should be completed by the relations following
from the q-deformed orthogonality condition.
\proclaim Proposition 7.5.
For the series $B,\ C$ and $D$, the matrix $Z$ fulfills also
$$
\delta_{jk}=\sum_s (Z_2C_2QZ_2^tQ^{-1}C_2^{-1})_{kj,ss}\, .
\eqno(7.14) $$

\noindent{\em Proof.}
Since $C\, L^tC^{-1}=L^{-1},\ C\, A^tC^{-1}=A^{-1}$, we have
$$
C(AZ)^tC^{-1}=Z^{-1}A^{-1}=Z^{-1}C\,A^tC^{-1}\, .
\eqno(7.15) $$
Furthermore, multiplying (7.12) by $C_2^{\ -1}$ from the left and by $C_2$
from the right one obtains
$$
A_2^{\ t}Z_1(A_2^t)^{-1}=\tilde Q^{-1}Z_1\tilde Q\, ,\quad
\hbox{where}\quad \tilde Q=C_2^{\ -1}Q\, C_2\, .
$$
Using this relation one can derive for the matrix elements
\begin{eqnarray*}
[(AZ)^t(A^t)^{-1}]_{jk} & = & \sum_s[A_2^tZ_1(A_2^t)^{-1}]_{ss,jk} \cr\cr
& = & \sum_s (\tilde Q^{-1}Z_1\tilde Q)_{ss,jk} \cr\cr
\end{eqnarray*}
Consequently,
\begin{eqnarray*}
[ZC(AZ)^t(A^t)^{-1}C^{-1}]_{jk} & = & \sum_{stu}
(ZC)_{js}[\tilde Q^tZ_1^t(\tilde Q^t)^{-1}]_{st,uu}(C^{-1})_{tk}\cr\cr
& = & \sum_s (Z_2C_2QZ_2^tQ^{-1}C_2^{-1})_{kj,ss}\, .\cr\cr
\end{eqnarray*}
In view of (7.15) we have arrived at the sought relations. \QED

In the generic case ($\Pi_0=\emptyset$) the dressing orbit is nothing
but the flag manifold. In this case $Q_{12}=Q_{21}=\diag R$ and $Z$ is an
upper triangular matrix with units on the diagonal. The relation (7.13) can
be simplified since $\diag R$ commutes with $R$,
$$
R\, Z_1\diago(R)\, Z_2=Z_2\diago(R)\, Z_1R\, .
\eqno(7.16) $$

For the series $A$, i.e., $K=SU_q(N)$ we have
\begin{eqnarray*}
R_{jk.st} & = & \delta_{js}\delta_{kt}+(q-q^{\sgn(k-j)})\,\delta_{jt}
\delta_{ks}\, ,   \cr\cr
Q_{jk.st} & = & q^{\delta_{jk}}\delta_{js}\delta_{kt}\, ,  \cr\cr
\end{eqnarray*}
and the relation (7.16) can be rewritten for the individual matrix
entries as
$$
q^{\delta_{ks}}z_{js}z_{kt}-q^{\delta_{jt}}z_{kt}z_{js}=
(q^{\sgn(k-j)}-q^{\sgn(s-t)})q^{\delta_{js}}z_{ks}z_{jt}\, .
\eqno(7.17) $$
The relations (7.17) are already known
\cite{T-T, Awata et al}. Originally they were obtained by expressing
the entries $z_{jk}$ by means of the q-minors ($j<k$),
$$
z_{jk}=\left | T^{1\dots j}_{1\dots j}\right |^{-1}_q
\left | T^{1\dots ..j}_{1\dots j-1,k}\right |_q\, .
$$
But this derivation seems to be rather tedious and doesn't suggest
the compact form (7.16).

\section{Representation acting in a space of antiholomorphic
functions}

Let us denote by \Cl the algebra of quantum holomorphic functions living on
the cell. This means that \Cl is generated by the entries of $Z$ fulfilling
(7.13) and possibly also relations following from the deformed orthogonality
condition. $\Cla^\ast$ stands for the algebra of antiholomorphic functions
determined by the adjoint relations. We know that every vector
$u\in\Hla$ is represented by an element $\psi_u\equiv\psi_u(z^\ast)$ from
$\Cla^\ast$
(c.f. (6.1)), the mapping $u\mapsto\psi_u$ is linear and injective
and the lowest weight vector is sent to the unit. Denote by
$\Mla\subset\Cla^\ast$
the image of \Hl. We wish to transcribe the representation \tl as acting in
\Ml, but without introducing a special symbol for this new
realization. We
recall that both $\Aq(K)$ and $\Aq(K_0\backslash K)$ become left
$\Uh$(\gtk)--modules provided one relates to every element
$Y\in\Uh$(\gtk)
the left--invariant map $\xi_Y$ on $K_q$,
$$
\xi_Y\cdot f=(\id\otimes\langle Y,\cdot\rangle)\,\Delta f\, ,\quad
f\in\Aq(K)\, .
\eqno(8.1) $$
Then $\Cla^\ast$ becomes a left $\Uh$(\gtk)--module with respect to
the action
$$
(Y,f)\mapsto\wla^{\ -1}\,\xi_Y\cdot(\wla f)\, .
\eqno(8.2) $$

\proclaim Proposition 8.1.
\Ml is the cyclic $\Uh$(\gtk)--submodule in $\Cla^\ast$
with the cyclic vector $1$, i.e.,
$$
\tla(Y)\psi=\wla^{\ -1}\,\xi_Y\cdot(\wla \psi)\, ,\quad
\mbox{for} \quad Y\in\Uh(\hbox{\gtk}),\ \psi\in\Mla, .
\eqno(8.3) $$

\noindent{\em Proof.}
The proof is done by the following chain of equalities,
\begin{eqnarray*}
\wla\,\tla(Y)\psi_u & = & \langle\Gamma,\tla(Y)u\rangle=
(\langle\ela,\tla(\cdot)\tla(Y)u\rangle\otimes\id)\rho\cr\cr
& = & (\langle\ela,\tla(\cdot)u\rangle\otimes\id\otimes
\langle Y,\cdot\rangle)
\rho_{12}\rho_{13}\cr\cr
& = & (\id\otimes\langle Y,\cdot\rangle)\,\Delta\,
(\langle\ela,\tla(\cdot)u\rangle\otimes\id)\,\rho\, .\cr\cr
\end{eqnarray*}
In the third equality we have used the identity
$$
\tla(Y)=(\tla(\cdot)\otimes\langle Y,\cdot\rangle)\rho\, .\ \QED
$$

Finally we are going to show that the reproducing kernel can be
introduced
also in the quantum case and the scalar product in \Ml can be
expressed with its
help. Let $\eta$ designates the Haar measure on $\Aq(K)$. We have the
orthogonality relations
\cite{W}
$$
\eta(\langle u_1,\Tla v_1\rangle^\ast\,\langle u_2,\Tla v_2\rangle)=
\Mrla^{\ -1}\,\langle v_1,v_2\rangle
\langle u_1,\tla(\gamma^{-1})\, u_2\rangle\, ,
\eqno(8.4) $$
where
$$
\gamma=\exp\Bigl(-{h\over 2}\,\sum_{\alpha>0} H_\alpha\Bigr)\, ,\
\Mrla={\rm tr}\,\tla(\gamma^2)\, .
$$
Letting $u_1=u_2=\ela$ in (8.4) we get
$$
\langle u,v\rangle=\cla\,\eta(\langle\ela,\Tla u\rangle^\ast
\langle\ela,\Tla v\rangle)\, ,
\eqno(8.5) $$
where $\cla=\Mrla\,\langle\ela,\tla(\gamma^{-1})\ela\rangle$.
Consequently,
$$
\langle u,v\rangle=\cla\,\eta(\psi_u^{\ \ast}\wla^{\
\ast}\wla\psi_v)\, .
\eqno(8.6) $$

Set now
$$
\Psi(z)=\Gamma\,(\wla^{\,\ast})^{-1}\in\Hla\otimes\Cla\, ,
\eqno(8.7) $$
and define the reproducing kernel as
$$
K(\zeta^\ast,z):=\langle\Psi(\zeta),\Psi(z)\rangle\in
\Cla^{\ \ast}\otimes\Cla\, .
\eqno(8.8) $$
Here $\zeta^\ast$ stands for the generators in $\Cla^{\ \ast}$ and
$z$ for those in \Cl.

It holds
$$
\langle u,v\rangle=\cla\,
\eta(\psi_u(z^\ast)^\ast K\bigl(z^\ast,z)^{-1}\psi_v(z^\ast)\bigr)\, .
\eqno(8.9) $$
It is enough to notice that $K(z^\ast,z)\in\Cla^{\ \ast}\cdot\Cla$ is
equal to $(\wla^{\ \ast}\wla)^{-1}$,
$$
K(z^\ast,z)=\wla^{\ -1}\,
\langle\ela,\Tla(\Tla)^{-1}\ela\rangle\,(\wla^{\,\ast})^{-1}=
\wla^{\ -1}(\wla^{\,\ast})^{-1}\, .
$$

Furthermore, substituting $\Psi(\zeta)$ for $u$ in (8.9) we obtain
$$
\psi(\zeta^\ast)=\cla\,\eta_z\bigl(K(\zeta^\ast,z)\, K(z^\ast,z)^{-1}
\psi(z^\ast)\bigr)\, ,\ \mbox{for every}\ \psi\in\Cla^{\ \ast}\, .
\eqno(8.10) $$

\section{Representations and non-commutative differential geometry}

We shall use the summation rule through this Section. All indices are
running form $1$ to $N$, $N$ being the dimension of the vector
representation. With some abuse of notation
we shall no more distinguish between the element $X\in \Uh$({\gt k}) and the
corresponding left--invariant mapping $\xi_X$ (8.1). We keep only the $\cdot$
to indicate the action of $\Uh$({\gt k}) on $\Aq(K)$. The
following notions and facts concerning the differential calculus on $\Aq(K)$
will be useful \cite{Wordc,Jurco, SW}. Let us denote as $M_{ijkl}$ the
following family of quantum functions on $K$
$$
M_{ijkl}=S^{-1}(U_{lj})U_{ik}.
$$
Let also
$$
f_{ijkl}= S^{2}(L^+_{jl})S(L^-_{ki})
$$
be a family of elements of $\Uh$({\gt k}).
We shall denote by ${\cal E}$ the free left module over
$\Aq(K)$ with generators denoted by $\Omega_{ij}$.
Let us introduce the right multiplication,
the right coaction $\delta_R$ and the left coaction $\delta_L$ of $\Aq(K)$
on ${\cal E}$ by
$$
a_{ij} \Omega_{ij}b = a_{ij} f_{ijkl}\cdot b \Omega_{kl},
$$
$$
\delta_R(a_{ij}\Omega_{ij})=\Delta a_{ij} (\Omega_{kl}\otimes M_{klij}),
$$
$$
\delta_L(a_{ij}\Omega_{ij})=\Delta a_{ij}(1\otimes \Omega_{ij}),
$$
for $a_{ij},\; b\in\Aq(K)$.
Then the triple $({\cal E},\delta_R, \delta_L)$ is an
$\Aq(K)$--bicovariant bimodule in the sense of \cite{Wordc}.
If we introduce quantum functionals $\chi_{ij} \in \Uh$({\gt k}) by
$$
\chi_{ij}=\delta_{ij} - L^-_{im}S(L^+_{mj}), \eqno(9.1)
$$
then the mapping $d:\;\Aq(K)\rightarrow {\cal E}$
$$
da = \Omega_{ij} \chi_{ij}\cdot a, \hskip 1cm a\in \Aq(K) \eqno(9.2)
$$
defines a bicovarint first--order differential calculus on $\Aq(K)$,
which extends uniquely
to the exterior differential calculus on $\Aq(K)$. The linear space
$\,_{inv}{\cal E}$
spanned by $\Omega_{ij}$'s is the space of left invariant one--forms.
Let us denote as $\,_{inv}{\cal X}$ the dual linear space of
left--invariant vector fields spanned by $\chi_{ij}$'s.
The linear space $\,_{inv}{\cal X}$ is closed under the q-commutator
$$
[X,Y]_q={\rm Ad}_X\, Y
$$
and the comultiplication on $\chi_{ij}$ reads
$$
\Delta \chi_{ij}=\chi_{ij} \otimes 1 + O_{ijkl}\otimes \chi_{kl},
$$
$$
O_{ijkl}=L^-_{ik}S(L^+_{lj}). \eqno(9.3)
$$

In the following we shall use freely the Cartan calculus on quantum groups
developed in \cite{Schupp, Aschieri}, where the inner derivation
$\hbox{\gt i}_{\xi}$ and the Lie derivative ${\cal L}_{\xi}$ of a general
$n$--form along a general vector field $\xi$ have been introduced.
Let us mention, without going into details, that the linear space of
left--invariant vector fields $\,_{inv}{\cal X}$ can be used to freely
generate an $\Aq(K)$--bicovariant bimodule ${\cal X}$ of general vector
fields on $\Aq(K)$. The right coaction of $\Aq(K)$ then coincides on
$\,_{inv}{\cal X}$ with the right dressing action ${\cal R}$ (3.3).
We refer the reader for more information to the above
mentioned papers.  Here we follow the conventions of \cite{Schupp}.

For any quantum function $a\in \Aq(K)$ let us introduce
the left--invariant one--form
$$
\Theta^a_L = da^{(2)}S^{-1}(a^{(1)})=\Omega_{ij}\langle \chi_{ij},a\rangle
\eqno(9.4)
$$
as well as the right invariant form
$$
\Theta^a_R = da^{(1)}S(a^{(2)})=\Theta^{a^{(2)}}_La^{(1)}S(a^{(3)})=
\Omega_{kl}S(U_{li})U_{jk}\langle \chi_{ij},a\rangle. \eqno(9.5)
$$

Let $X \in \,_{inv}{\cal X}$, then
we have for its symbol $\sigma(X)$
$$
\sigma(X)= \hbox{\gt i}_X \Theta_R^{\wla}. \eqno (9.6)
$$
This equality is a consequence of
a chain of identities which employs the rules of \cite{Schupp}
$$
\sigma(X)=\langle X,\wla^{(2)} \rangle \wla^{(1)}S(\wla^{(3)})=
(\hbox{\gt i}_X
\Theta_L^{\wla^{(2)}}) \wla^{(1)}S(\wla^{(3)}) = \hbox{\gt i}_X
\Theta_R^{\wla}.
$$
Applying the differential $d$ to the equality (9.6), making use of the
identity
$$
{\cal L}_X =\hbox{\gt i}_Xd+d\hbox{\gt i}_X,
$$
which remains to be valid also in the quantum case and using the fact that
$$
{\cal L}_X\omega = 0
$$
for $X$ left--invariant and $\omega$ right--invariant we obtain immediately
$$
d \sigma(X) =-\hbox{\gt i}_Xd\Theta_R^{\wla}.\eqno(9.7)
$$
Let $Y\in {\cal X}$ be now another left--invariant vector field.
We have
$$
-\hbox{\gt i}_Y\hbox{\gt i}_Xd\Theta_R^{\wla}= \sigma([Y,X]_q),\eqno(9.10)
$$
which follows from an application of $\hbox{\gt i}_Y$ to the equality (9.7):
$$
-\hbox{\gt i}_Y\hbox{\gt i}_Xd\Theta_R^{\wla}=\hbox{\gt i}_Yd \sigma(X)=
Y\cdot \, \sigma(X)= \sigma({\rm Ad}_Y\,X).
$$
The third equality in the above chain is a direct consequence of
definitions adopted in \cite{Schupp}.

Here we would like note the following.
Let us assume the image $\sigma(\Uh$({\gt k})$\subset \Aq(K_0\backslash K)$
under symbol mapping $\sigma$ equipped with a new product $\ast$, which
respects the algebra structure of $\Uh({\gt k})$
$$
\sigma (X)\ast \sigma (Y)=\sigma(XY) \hskip 1cm\mbox{for}\, X,\,Y\in
\Uh({\gt k}),
$$
which is just the Berezin quantization prescription for the symbols
in the classical case.
Then from (5.15) it follows immediately that the mapping $\sigma$ is a
quantum momentum map in the sense of \cite{Lu} and we can rewrite (9.10) in
the form
$$
-\hbox{\gt i}_Y\hbox{\gt i}_Xd\Theta_R^{\wla}= \sigma(Y^{(1)})\ast
\sigma(X)\ast \sigma(S(Y^{(2)})).
\eqno(9.11) $$

Using the expression of the right invariant form $\Theta_R^{\wla}$
with the help of the left
invariant forms $\Omega_{ij}$ following from (9.5) we obtain an alternative
definition of the isotropy subgroup $K_0$. Instead of (5.4) we may equivalently
require the invariance of $\Theta_R^{\wla}$ with respect to the left
coaction of $K_0$
$$
(p_0\otimes id)\delta_L \Theta_R^{\wla}=1\otimes \Theta_R^{\wla}.\eqno(9.12)
$$

Let us denote for convenience by ${\cal Z}\in \Hla\otimes\Aq(K_0\backslash K)$
the unnormalized coherent state ${\cal Z}=\Gamma (\wla^{\ast})^{-1}$ and let
the expressions $d{\cal Z}$ and $d\Gamma$ have the obvious meaning of
differentiating with respect to the second factor in $\Hla\otimes\Aq(K)$.
Let us also introduce a new one--form $\Theta^{\wla}\equiv
\Theta_R^{\wla}-d\wla (\wla)^{-1}$.
Like in the classical case the one--forms $\Theta_R^{\wla}$ and
$\Theta^{\wla}$ can be expressed through the coherent states $\Gamma$ and
${\cal Z}$ as
$$
\Theta_R^{\wla}=\langle d\Gamma|\Gamma\rangle
$$
and
$$
\Theta^{\wla}=\wla \langle d{\cal Z}|{\cal Z}\rangle \wla^{\ast},
$$
respectively.

Now we are prepared to give
a formula for the action of the elements $\chi_{ij}$ in the irreducible
$\ast$--representation \tl of $\Uh$({\gt k}), which directly generalizes
the geometric quantization prescription for the action of generators of
$U$({\gtk}) in the  irreducible representation of $K$ corresponding
to a minimal weight $\lambda$.
Starting from formula (8.3) and using (9.3) we have
$$
\tla(\chi_{ij})\psi=\wla^{-1} (\chi_{ij}\cdot \wla)\psi +
\wla^{-1}(O_{ijkl}\cdot \wla)
\chi_{kl}\cdot \psi,
$$
which can be finally rewritten making use of the following identities
$$
(\chi_{ij}\cdot \wla)\wla^{-1}= \hbox{\gt i}_{\chi_{ij}}d\wla \,
\wla^{-1}=\sigma(\chi_{ij})-
\hbox{\gt i}_{\chi_{ij}} \Theta^{\wla}
$$
in the form
$$\tla(\chi_{ij})\psi=\wla^{-1}(O_{ijkl}\cdot \wla)
\chi_{kl}\cdot \psi +\wla^{-1}(\sigma(\chi_{ij})-
\hbox{\gt i}_{\chi_{ij}} \Theta^{\wla})\wla \psi.\eqno(9.13)
$$
\vskip 12pt
{\bf Acknowledgements.}\
B.J. would like to thank  D. Arnaudon, P. Aschieri, M. Bauer,
L. Castellani, M. Dijkhuizen, T.H. Koornwinder, S. Majid and P. Truini
for helpful discussions.  P.S. wishes to acknowledge gratefully partial
support from CGA grant 202/93/1314.

\end{document}